\documentclass{aa}
\usepackage{txfonts}
\usepackage{latexsym}
\usepackage[latin1]{inputenc}
\usepackage{exscale}
\usepackage[below]{placeins}
\usepackage{graphicx}
\usepackage{amsfonts}
\usepackage{graphicx,epsfig}
\usepackage{natbib}
\usepackage{aas_macros}

\begin{document}
\title{Automatic detection of arcs and arclets formed by gravitational lensing}

\author{Frank Lenzen\inst{1}, Sabine Schindler\inst{2} and Otmar
        Scherzer\inst{1}}
\authorrunning{F. Lenzen \and S. Schindler and O. Scherzer}
\titlerunning{Automatic detection of arcs and arclets}
\institute{Institute of Computer Science, University of Innsbruck,
Technikerstra\ss e 25, 6020 Innsbruck, Austria
\and Institute for Astrophysics, University of Innsbruck,
Technikerstra\ss e 25, 6020 Innsbruck, Austria }
\offprints{Frank Lenzen, \email{Frank.Lenzen@uibk.ac.at}}
\date{Received ... / Accepted ...}

\abstract{We present an algorithm developed particularly 
to detect gravitationally lensed arcs
in clusters of galaxies. This algorithm is suited for automated
surveys as well as individual arc detections. New methods are used for
image smoothing and source detection. The smoothing is performed by so-called
anisotropic diffusion, which maintains the shape of the arcs and does
not disperse them. The algorithm is much more efficient in detecting
arcs than other source finding algorithms and the detection by eye.
}
\maketitle

\section{Introduction}

Gravitational lensing has turned out to be a universal tool for very
different astrophysical applications. In particular,
lensing by massive clusters of galaxies is extremely
useful for cosmology. The measurement of
various properties of the magnified and distorted images of background
galaxies (``arcs and arclets'') provides information on
the cluster as well as on the population of faint and distant
galaxies. 
Many of these distant background galaxies (up to redshifts of $z \approx 5$ 
, \cite{FRANX97})
could not be studied with 
the largest telescopes if they were not magnified by the gravitational 
lensing effect.  Some of these distant galaxies are
particularly useful for the study of galaxy evolution
\cite[]{SEITZ98,PETTINI00}.
As these background galaxies are free of selection effects, because they lie
serendipitously behind massive clusters, they are ideal
targets for a population study of distant galaxies \cite[]{FORT97}.
Gravitationally lensed arcs also provide a way to measure the
total mass and the dark matter in clusters \cite[]{FORT94,
WAMBSGANSS98, Mellier99}. As
galaxy clusters
can be considered to be fair samples of the universal mass fractions, such 
determinations probe 
cosmological parameters
like $\Omega_{\rm tot}$,  $\Omega_{\rm matter}$,  and $\Omega_{\rm baryon}$. 
A third, very important application of gravitational lensing in clusters
is the determination of
the frequency of arcs (=arc statistics). This 
is a strong criterion in order to distinguish
between different cosmological models \cite[]{BARTELMANN98,
KAUFMANN00}. Therefore detections of gravitationally lensed 
arcs are of high importance for astrophysics and cosmology.

Ideal for cosmological studies are systematic searches.
A successful arc search was
performed with the X-ray luminous cluster 
sample of the EMSS \cite[]{LUPPINO99}. More searches are
under way, which not only cover larger areas 
than the previous survey, but
they also go much deeper, i.e. fainter galaxies can be
detected.

The first arcs were detected only in 1986 \cite[]{LYNDS86, SOUCAIL87}
 because they are very faint and very thin structures. Under
non-ideal observing conditions (e.g. bad seeing) 
they are easily dispersed and disappear
into the background. Even with ideal conditions they are not easy to
detect because they are often just above the background level. In
 order to
remove the noise and make faint structures better visible usually
smoothing is applied. Unfortunately in the case of such thin
structures as arcs the smoothing procedure often leads to a dispersion of
the few photons so that the arcs are difficult to detect at all.
To prevent this dispersing we suggest an algorithm that
automatically smooths only along the arcs and not perpendicular to
them, so-called ``anisotropic diffusion''. The subsequently applied
source finding procedure extracts all the information from the sources
necessary to distinguish arcs from other sources (i.e. edge-on
 spirals or highly elliptical galaxies). This new algorithm
is much more efficient in finding gravitationally lensed arcs than existing
 source detection algorithms,  because it is
optimized just for this purpose. 

In Sect.~2 the algorithm is explained with its four different steps.
In Sect.~3 examples of detected arcs are presented.
Sect.~4 outlines the differences to existing source
finding software and the advantages for arc detection. In Sect.~5 we draw
conclusions on the applicability and usefulness of the new algorithm.

\section{The algorithm}

We propose a \emph{four level} strategy for numerical detection of arcs
in astronomical data sets consisting in the successive realization of
\begin{enumerate}
\item histogram modification \item anisotropic diffusion filtering
\item object finding \item selection of arcs.
\end{enumerate}
The algorithm described in detail below has been implemented 
in the programming language C.\\

The \emph{image data} are given by a 2D matrix of 
\emph{intensity values} $I$. For the sake of simplicity of presentation
we assume that the intensity matrix is of dimension $N \times N$.
The set of indexes
\[P:=\{(i,j),\;\,i,j=1\dots N\}\]
are referred to as pixels and identify
the position of the recorded intensity $I(i,j)$.

\subsection{Histogram modification}
\label{subsec:HistogramMod}
Astronomical image data contain objects on a variety of
brightness scales. Frequently stars and galaxies show up
relatively bright; arcs however are small \emph{elongated} objects of 
only marginally 
higher intensity than the surrounding
background. In order to detect such arcs 
it is necessary to correct for the dominance of extremely
bright objects. This is done by \emph{histogram modification}.
Here, we use a nonlinear transformation
\begin{equation}
\label{scaling}
\mathcal{S}(x)=\left\{
\begin{array}{cl}
0&\mbox{ if }x<a\\
s(\frac{x-a}{b-a})&\mbox{ if } x\in [a,b]\\
1&\mbox{ if } x>b
\end{array}
\right.
\end{equation}
which is applied to the pixel intensities $I(i,j)$, giving a new intensity
matrix $I^0(i,j)=\mathcal{S}(I(i,j))$. The transformation $\mathcal{S}$ 
maps the pixel
intensity distribution, which originally varied from 0 to the maximum
pixel value, onto the interval  $[0,1]$ by the use of a bijective 
transformation $s$ from the interval $[0,1]$ to $[0,1]$.

\begin{figure}[h]\begin{center}
\includegraphics[width=0.45\textwidth]{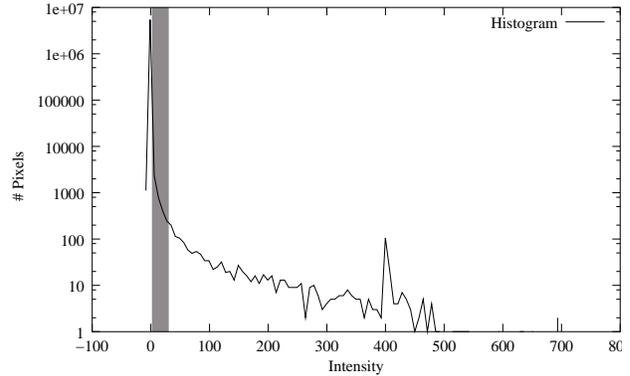}
\caption{\label{FigHist1} Distribution of pixel values 
of an astronomical test image.
A typical parameter setting for $a$ and $b$ 
(low and high cut) is marked in gray. 
A good choice for parameter $a$  
is the maximum of the distribution and can be easily computed ,
 whereas a general optimum for parameter $b$ can not be prescribed.
In the considered example a good choice is: $a=0$ and $b=1$.
}
\end{center}\end{figure}

The interval $[a,b]$ specifies the level of intensities where
arcs are to be detected. The lower bound $a$ is considered
the intensity value of the background. By
analyzing several different astronomical data sets we have learned that $a$ and
$b$ have to be chosen relatively close to each other for optimal
visualization of arcs (cf. Fig.~\ref{FigHist1}).

In the interval $[a,b]$ 
we have to distinguish between noise and real sources such as stars, 
galaxies and arcs. 
To ease this separation process we apply nonlinear intensity transformations 
such as $s(x)= \sqrt{x}$, or alternatively $s(x)=x$.

Some astronomical data set may contain arcs in different
intensity ranges. In this case choosing the value for the parameter 
$b$ of about the intensity of the brightest arcs is appropriate.

\subsection{Anisotropic Diffusion Filtering}
\subsubsection{Introduction to diffusion processes}
By interpolation the scaled image $I^0(i,j)$ can by interpreted as a function
in $\mathbb{R}^2$ and is now denoted by $u^0(x,y),\;(x,y)\in \mathbb{R}^2$.

By applying Gaussian convolution with a kernel 
$K_t(x,y):=\frac{1}{4\pi t}\exp \left(-\frac{x^2+y^2}{4t} \right)$ depending on $t$ one 
gets a smoothed image 
\begin{eqnarray*}
u(t,x,y)&:=&u^0(x,y)\ast K_t (x,y)\\
         &:=&\int_{\mathbb{R}^2} u^0(x-\tilde{x},y-\tilde{y})\;K_t(\tilde{x},\tilde{y})\;d\tilde{x}d\tilde{y}.
\end{eqnarray*}
The parameter $t$ controls the amount of smoothing. 
A larger value of $t$ corresponds to a higher filtering.

In the following $\partial_t u(t,x,y)$ denotes the first derivative of $u$ with respect to $t$,
$\partial_{xx} u(t,x,y)$ and $\partial_{yy} u(t,x,y)$ the second derivatives with respect to $x$ and $y$
and $\Delta u(t,x,y):=\partial_{xx} u(t,x,y)+\partial_{yy} u(t,x,y)$ the Laplacian. 

It is well known that $u(t,x,y)$ solves the diffusion (resp. heat) equation
\begin{equation}
\partial_t u(t,x,y) -\Delta u(t,x,y)=0 \label{HeatEq}\mbox{ for all } (x,y)\in\mathbb{R}^2
\end{equation}
with the initial condition
\[
u(0,x,y)=u^0(x,y)
\]

To simplify the notation we write $u(t,x),\;x\in\mathbb{R}^2$ instead of $u(t,x,y)$.\\

Eq. (\ref{HeatEq})  can be restricted to a rectangular domain $\Omega\subset \mathbb{R}^2$, the domain of interest,
typically the set of pixels where intensity information has been recorded.
 
In order to achieve existence and uniqueness of the solution $u(t,x),\;x\in\Omega$ 
it is necessary to prescribe boundary conditions such as
\[
\frac{\partial u}{\partial n}(t,x) = 0 \mbox{ for all } t\in(0,\infty) \mbox{ and } x\in\partial\Omega,
\]
where $\partial \Omega$ denotes the boundary of the domain $\Omega$, $n$ is the outer normal vector on $\partial \Omega$ 
and $\frac{\partial}{\partial n}$ denotes the derivative in direction of $n$.

Applying the diffusion process up to a fixed time $T>0$ smooths the given data $u^0$
and spurious noise is filtered. The parameter $T$ defines the strength of the filtering  process. 
Thus in the following we refer to $T$ as the \emph{filter parameter}.\\

The disadvantage of Gaussian convolution is that edges in the filtered image $u(T,x)$ are blurred and the allocation and detection of object borders is difficult. To settle this problem several advanced  diffusion models have been proposed in the literature \cite[]{WEICKERT,CLMC}.

In the next section we define the general model of a diffusion process and in Sect.~\ref{subsubsec:SpecificModel} we describe the specific model used in our algorithm.

\subsubsection{The general diffusion equation}

\emph{Anisotropic diffusion filtering} consists in 
solving the time
dependent differential equation
\begin{equation}
\label{eq:anis}
\partial_t u(t,x) - \mbox{div}\Big(D(x,u,\nabla u)\nabla u(t,x)\Big) =0
\end{equation}
up to a certain time $T > 0$.

Here $D(x,u,\nabla u)$
is the $2\times 2$ \emph{diffusion matrix} depending on $x$ and $u$,
 
We prescribe the same initial and similar boundary conditions as mentioned above.
Setting $D(x,u,\nabla u)=1$ results in the heat equation (\ref{HeatEq}).

Two classes of anisotropic diffusion models are considered in the
literature: if $D(x,u,\nabla u)$ is independent of $u$ and $\nabla u$ then
Eq. (\ref{eq:anis}) is called \emph{linear} anisotropic diffusion filtering,
otherwise it is called \emph{nonlinear} filtering.
 
For a survey on the topic of diffusion filtering we refer to \cite{WEICKERT}.

\subsubsection{The specific model\label{subsubsec:SpecificModel}}

In anisotropic models the diffusion matrix $D$ is constructed to
reflect the estimated edge structure. That is to prefer smoothing in
directions along edges, or in other words edges are preserved or even enhanced
and simultaneously spurious noise is filtered.

Consequently this kind of filtering eases a subsequent edge-based 
object detection.

In the following $D$ will only depend on the gradient $\nabla u$, which reflects 
the edge structure of the image.

To accomplish the diffusion matrix $D(\nabla u)$ we note that
\[
      v_1:=\frac{\nabla u}{|\nabla u|}
      \mbox{ and }
      v_2:=\frac{\nabla u^\bot}{|\nabla u|} :=
      \frac{1}{|\nabla u|}
      \left( \frac{\partial u}{\partial y},
      -\frac{\partial u}{\partial x} \right)
\]
denote the directions perpendicular and  parallel, respectively, to
edges. (By 
$v^\bot=\left(
	\begin{array}{c}
	v_1\\
	v_2\\
	\end{array}\right)^\bot 
	=\left(
	\begin{array}{c}
	v_2\\
	-v_1\\
	\end{array}\right)
	$ 
	we denote the vector perpendicular to $v$).
     
By selecting
\[
      D(\nabla u) = (v_1,v_2) \left(\begin{array}{cc}
                                    g(|\nabla u|) & 0\\
                                    0 & 1
                    \end{array} \right) (v_1,v_2)^T\;.
\]
with
\[
	g(|\nabla u|):=\frac{1}{1+\left(\frac{|\nabla u|}{K}\right)^2}\,,
	\quad \mbox{ with parameter }K>0 
\]
the diffusion filtering method (\ref{eq:anis}) prefers filtering
parallel to edges.

\begin{figure}[h]\begin{center}
\includegraphics[width=0.45\textwidth]{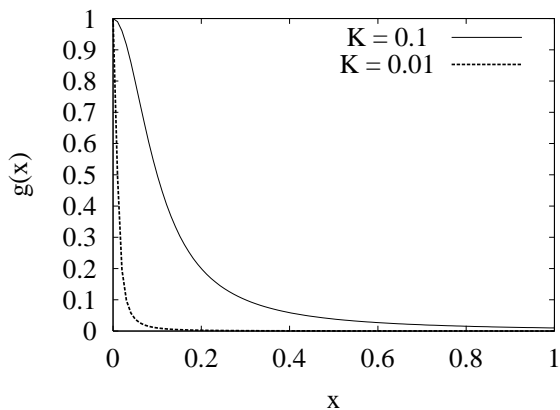}
\caption{\label{FigFunctionG}
Graph plot of function g(x) used for weakening the diffusion orthogonal 
to edges to archive edge enhancing. For parameter K 
values $K=0.1$ resp. $K=0.001$ are used.
}
\end{center}
\end{figure}

\begin{figure}[h]\begin{center}
\includegraphics[width=0.45\textwidth]{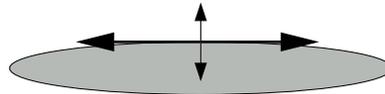}
\caption{\label{FigDiffusion} 
Diffusion (smoothing) near edges: The thickness of the arrows indicate the strength of diffusion. Parallel to the edge a strong diffusion occurs whereas in orthogonal direction a weak diffusion leads to an enhancing of the edge. Due to the averaging of the structure tensor these directions are also determine the diffusion in a surrounding area of the edge and in particular at the vertices yielding a diffusion mainly parallel to the direction of elongation. 
}
\end{center}
\end{figure}

Fig.~\ref{FigDiffusion} highlights the diffusion directions:
arrows indicate the main directions of diffusion ($v_1,v_2$) and their
thickness relates to the diffusion coefficient determining the 
strength of diffusion. Parallel to the edge the diffusion 
coefficient is constantly 1 (strong diffusion), where as the 
diffusion coefficient $g(|\nabla u|)$ in normal direction
decreases rapidly (c.f. Figure~\ref{FigFunctionG})
as $|\nabla u|$ increases (weak diffusion on edges).

The dependence of $g(|\nabla u|)$ on $|\nabla u|$ is 
controlled by parameter $K$ (c.f. Fig.~\ref{FigFunctionG}). 
We therefore refer to $K$ as the \emph{edge sensitivity} parameter.\\

We use the following common variation of the diffusion 
matrix $D$:

To limit the effect of noise the diffusion tensor 
is chosen to be dependent on the pre-smoothed image 
$u_\sigma(t,x)=u(t,x)\ast K_\sigma$
obtained by Gaussian convolution with \emph{pre-filter parameter} 
$\sigma>0$. In the following to determine the diffusion matrix
we exploit the gradient of of the pre-filtered image $u_\sigma(T,.)$  
instead of $u(T,.)$.

Let $v_1^\mu,v_2^\mu$ be the eigenvalues of the filtered
structure tensor
\[
        J_\mu(x):=\left(K_\mu *
        \left[\left(\nabla u_\sigma\right)
        \left(\nabla u_\sigma\right)^T\right]\right)(x)\;.
\]
We refer to $\mu\ge 0$ as the 
\emph{pre-filter parameter for the structure tensor}.

Note that in the case $\mu=0$ the eigenvectors of $J_0$ are
$\frac{\nabla u_\sigma}{|\nabla u_\sigma|}$ and
$\frac{\nabla u_\sigma^\bot}{|\nabla u_\sigma|}$.
If $\sigma$ and $\mu$ are small, the effect of Gaussian filtering
is negligible, and consequently $v_1^\mu, v_2^\mu$
and $v_1,v_2$ refer to similar edge structures.

The purpose of filtering of the structure tensor is to
average information about edges in the image.\\

Taking into account the approximation we are led to the following
diffusion matrix
\[
      D_{\mu,\sigma}(\nabla u_\sigma) = (v_1^\mu,v_2^\mu) \left(\begin{array}{cc}
                                    g(|\nabla u_\sigma|^2) & 0\\
                                    0 & 1
                    \end{array} \right) (v_1^\mu,v_2^\mu)^T\;.
\]

Besides the fact that $D_{\mu,\sigma}(\nabla u_\sigma)$ is less noise sensitive there are several advantages of using the approximation
$D_{\mu,\sigma}(\nabla u_\sigma)$ instead of $D(\nabla u)$:
\begin{enumerate}
\item The numerical calculation of $D(\nabla u)$ is unstable; if
            $\mu$ and $\sigma$ are chosen appropriately 
	    $D_{\mu,\sigma}(\nabla u_\sigma)$
            can be determined in a stable way.
\item For our particular application, in a neighborhood of arcs
            the vector field $v_2^\mu$ is nearly parallel to the
            arcs. Thus filtering is performed in arc orientation
            enhancing the geometrical structure of arcs. A side effect which 
            is very useful for our particular application is that small 
            gaps between elongated structures are closed, merging nearby 
            objects. 
\end{enumerate}

This is the anisotropic diffusion filtering method used in our
numerical calculations.

In order to solve the differential equation (\ref{eq:anis}) it is discretized
 using a \emph{finite element method} in space and an \emph{implicit Euler method} in time.
(Within the finite element method the width of the quadratic elements is set 
to 1.)
For each time step the resulting system of linear equations is solved by a
\emph{conjugate gradient method}.
For a survey on solving parabolic differential equations with finite element methods we refer to \cite{THOMEE}.
The conjugate gradient method is discussed in \cite{HANKE95,HANKE02}.

Anisotropic diffusion filtering requires to select
parameters $T,K,\sigma$ and $\mu$.
\begin{itemize}
\item \emph{Filter parameter} $T$ defines the amount of filtering.
      Consequently for small values of $T$, spurious noise is still
      recognizable. If $T$ is large, spurious noise is eliminated
      but also small details are blurred. The parameter $T$ has to be
      adapted to each data set to be analyzed.
\item \emph{Edge sensitivity parameter} $K$ determines 
      the dependency of the strength of diffusion orthogonal
      to edges on $|\nabla u_\sigma|$.
      The smaller $K$ is, the more edges are enhanced. 
      Thus, if $K$ is chosen too low, 
      the diffusion may generate artifacts out from the noise.
      Using a too large value for $K$ leads to a filtering similar to 
      Gaussian convolution, i.e. the image gets blurred.\\
      The parameter can be calibrated using test data and remains unchanged for similar data.
\item \emph{Pre-filter parameter} $\sigma$ can be selected from
      calibrated data.
\item The \emph{pre-filter parameter for the 
      structure tensor}  $\mu$ controls the local average information. 
      It has to be adopted to the size of the object to be recovered.
\end{itemize}

\subsection{Object finding}
In this section we discuss a \emph{partitioning} algorithm to
separate different image data, i.e. disjoint subsets of connected
objects and background (\emph{partitions}). 
The algorithm uses \emph{only} the anisotropic diffusion filtered data $u(\cdot):=u(T,\cdot)$
and not the initial data $u^0$.

In order to save computational effort we restrict our attention to segment objects of interest, i.e. 
isolated objects exceeding a certain brightness. 
We search for local intensity maxima exceeding a certain intensity $c_{\rm min}$ 
(referred to as the \emph{intensity threshold for detection}). 

Each maximum serves as \emph{seed} for the partitioning algorithm:
Starting from the seed pixel the region to which this pixel belongs has to be determined.

To outline this concept we use the following notation.
For a given pixel $p =(i,j)\in P$ we denote by 
$N(p)$
the set of the eight neighboring pixels.\\
The \emph{neighborhood} of a set $R \subseteq P$ is the set
\[ N(R)=\{q \in N(p) : p \in R\}\;.\]
For two pixels $p,q\in P$ a \emph{path connecting $p$ to $q$}
is any sequence $(p=p_0,p_1,p_2,\dots p_n=q)$ satisfying
$p_i \in N(p_{i-1})$, $i=1\dots n$.
\bigskip\par\noindent

The partitioning algorithm consists of two loops.
{\tt
\begin{enumerate}
\item Set $j=0$ and $P_0:=P$ for initialization.
\item \label{it1}
      Update $j \to j+1$.

      A pixel $p \in P_{j-1}$ is selected where the intensity of the filtered
      image $u(p)$ attains a local maximum exceeding the intensity 
      threshold for detection:  $u(p)>c_{\rm min}>0$.\\
      A numerical procedure for detection of
      local maxima is described in Sect.~\ref{subsubsec:detectionLocMax}.

      If no such pixel can be found the algorithm is terminated and
      $P_{j-1}$ is called ``background''.
\item \label{it2}
      The second step is a \emph{region growing algorithm}.
      \begin{enumerate}
      \item It is initialized with $i=0$,
            the \emph{seed} $p_0:=p$ and the region $R_0:=\{p_0\}$.
            For $p_0$ a threshold parameter $c_{\rm thres}>0$ is selected,
            which is used to terminate the region growing algorithm.
            The determination of this parameter is explained in
            Sect.~\ref{subsec:det}.
      \item Update $i \to i+1$.

            $R_i$ consists of $R_{i-1}$ and
            the neighboring pixels $N(R_{i-1})\cap P_{j-1}$ satisfying that
            \begin{itemize}
            \item the according intensity exceeds $c_{\rm thres}$ and
            \item the intensity is smaller than the intensities
                  of neighboring pixels in $R_{i-1}$.
            \end{itemize}
      \item The iteration is terminated when $R_{i+1}=R_i$ and
            we denote $R(p_0):=R(i)$. This region corresponds to an object.
      \end{enumerate}
\item Updating $P_{j}= P_{j-1} \backslash R(p_0)$ and
      repeating steps \ref{it1} and \ref{it2}, completely determines the
      algorithm.
\end{enumerate}
}

\begin{figure}[h]\begin{center}
\includegraphics[width=0.45\textwidth]{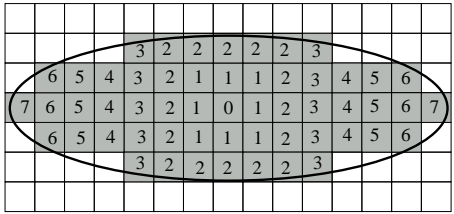}
\caption{\label{FigSegmentation}
Segmentation of an elliptical object (the theoretical object boundary is indicated by the black ellipse, gray squares indicate pixels with high intensities / white squares indicate pixels with low intensities): Assuming that the region growing starts with the pixel numbered by zero the neighboring pixels with numbers 1 to 7 are successively added to the region until the intensities of the next pixels to be added (here: white colored ) fall below a certain thres\-hold intensity.
}
\end{center}
\end{figure}

Fig.~\ref{FigSegmentation} illustrates the process of region growing for an elliptical objects.\\

A detailed overview over segmentation methods is given in
\cite{ROSENFELD1} or \cite{SOILLE}.

In the following we describe numerical procedures for calculation of local
maxima and the determination of $c_{\rm thres}$ in the object finding algorithm.

\subsubsection{Determination of $c_{\rm thres}$}
\label{subsec:det}

Let $p_0$ be a local maximum within an arc, which is used as an
initialization for the region growing algorithm.
The anisotropic diffusion filtering
method calculates the two eigenvalues $v_1^\mu$ and $v_2^\mu$ of
the structure tensor $J_\mu$. Thinking of an arc as an elliptically
shaped region $v_1^\mu$ and $v_2^\mu$ approximate the axes of the
ellipse. $v_1^\mu$, $v_2^\mu$ denote the principal, respectively
cross-sectional axis. Let $u_{-n},u_{-n+1},\dots u_0,\dots u_n$ be
the intensities along the cross-sectional axes.
\begin{figure}[h]\begin{center}
\includegraphics[width=0.45\textwidth]{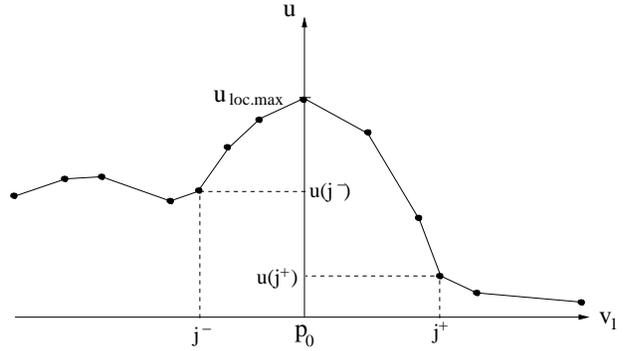}
\caption{\label{FigCrossSection}Typical intensity distribution along a 
cross-section through an object}
\end{center}\end{figure}
Fig.~\ref{FigCrossSection} shows such a typical intensity distribution 
along a cross-sectional axes of an object. The points on the cross sectional 
axis with maximal gradient are natural candidates for object boundaries. In the
discrete setting these points are
\begin{eqnarray*}
j^-&=&\arg \max_{i=-1\dots -n}\{|u_{i}-u_{i+1}|\}\,,\\
j^+&=&\arg \max_{i=1\dots n}\{|u_{i}-u_{i-1}|\}\;.
\end{eqnarray*}
As threshold intensity for the region growing algorithm we choose:
\[
c_{\rm thres}:=\max\{{u_{j^-},u_{j^+}}\}
\]

\subsubsection{Detection of local maxima}
\label{subsubsec:detectionLocMax}
For the detection of local maxima which exceed the \emph{intensity thres\-hold for detection}, $c_{\rm min}$,
 we proceed in a similar way as in the implementation of watershed algorithms (c.f. \cite{STOEV2000}). 
The strategy for finding a local maximum is to choose an initial pixel $p$ and to look for a pixel $q$ 
neighboring to $p$ with higher intensity and then with $p\leftarrow q$ reinitialized to proceed iteratively 
until a local maximum is reached.

Applying this procedure we have to deal with the case, that a local maximum may not be a single pixel 
but a connected set of pixels with the same intensity 
(a \emph {plateau}). 
 In the sequel, for simplicity of presentation, we also refer to a single pixel as a 
plateau.\\

Taking this into account we use the following procedure:\\

We use  markers (+), (-), and (0) to denote if a pixel \emph{is a local maximum}, \emph{not a local maximum}, 
 or if it \emph{is not yet considered} in the algorithm, respectively. Initially every pixel is marked by (0).

{\tt
\begin{enumerate}

\item
	Search for a pixel $p$ marked with (0). 
	If no such pixel exists, the algorithm is terminated.\\
	If the  intensity $u(p)$ lies below $c_{\rm min}$ mark $p$ with (-) 
	and repeat this step with another pixel.
\item
	Start segmentation of the plateau including pixel $p$ by applying 
	a region-growing algorithm. 
	During this process we monitor the following occurrences:
	\begin{enumerate}
	\item
		A pixel $q$ neighboring the plateau is found with $u(q)>u(p)$:
		Mark the pixels of the plateau with (-), set $p=q$ and repeat
		 step 2.
	\item
		A pixel of the plateau is found which is already marked with
		 (-):
		Mark the pixels of the plateau with (-) and go to step 1.
	\item
		All neighboring pixels of the plateau have less intensity:
		Mark the pixels of the plateau with (+) and go to step 1.
	\end{enumerate}
\end{enumerate}
}

After finding a local maximal plateau we choose one pixel
of the plateau as \emph{seed} for the later object detection.

Note that the identification of  local maximal plateaus can not be carried out by investigating 
the first and second derivatives of the intensity function.

Threshold $c_{\rm min}$ correlates to the parameters $a$ and $b$. The latter should be chosen such that $a$ is 
about the background intensity and $b$ is about the intensity of the arcs to be detected.   
As the arc intensities are only very little higher than the noise amplitude, $c_{\rm min}$ serves as a 
threshold between the noise level and arcs intensity range and is of high importance for the object detection.

Choosing $c_{\rm min} \approx 0$ guarantees detection of most objects
but many of them may result from noise amplification.
On the other hand a choice  $c_{\rm min}\approx 1$ real arcs with intensity below $c_{\rm min}$
are obviously not detected.\\

\subsection{Selection of Arc Candidates}
\subsubsection{Selection of Source Candidates}
\label{subsec:detectionCand}

Due to blurring effects on the CCD, in a neighborhood of bright stars and
galaxies, the background shows up bright, amplifying the intensity
of the noise. In these regions several local intensity maxima
above the threshold $c_{\rm min}$ occur and regions are detected,
which clearly are artificial and belong to the background. Such
regions are singled out by a comparison of the mean intensities along
a cross section within an object and the background. To this end
let
\begin{eqnarray*}
u_{\rm obj}&:=&\frac{1}{j+|j^-|+1}\sum_{i=j^-\dots j^+} u_i\\
u_{\rm back}&:=&\frac{1}{j+|j^-|}\left(
\sum_{i=2j^-\dots (j^--1)} u_i+\sum_{i=(j^++1)\dots 2j^+} u_i\right)
\end{eqnarray*}
The artificial objects (e.g. resulting from the diffusion filtering of noise) 
satisfy the condition that $u_{\rm obj}-u_{\rm back}$ is small.
Consequently we further restrict our attention to objects
satisfying
\[u_{\rm obj}-u_{\rm back}>c_{\rm min2} > 0\]
The parameter $c_{\rm min2}$ denotes the minimal difference of the objects 
intensity from the background.
We refer to it as the parameter of \emph{ minimal intensity difference from background}.

By taking into account only pixels with indices
between $2j^-$ and $2j^+$, $u_{\rm back}$ 
averages the intensities in a small neighborhood of the object.

We assume that the object is isolated, i.e. that there are 
no other objects in this neighborhood and 
$u_{\rm back}$ reflects the local average background intensity.

\subsubsection{Selection of Arcs}
\label{subsubsec:detection}

The last step of our algorithm consists in selecting arcs from
the detected objects. 
For deriving the following features of the objects the image 
resulting from applying the histogram modification is used.\\

Let $R$ be an object, consisting of the set of pixels
\[ \{ p_i=(x^1_i,x^2_i) : i\in N\} \subseteq P\]

The \emph{light intensity} of such an object can be interpreted
as \emph{mass} resp. \emph{density} and thus regarding the object as an rigid body we define
the total intensity
\[
m:=\sum_{i\in N} u^0(p_i)\;.
\]
The \emph{object's center } is given by
\[
mc=\left(
\begin{array}{c}
mc^1\\
mc^2
\end{array}
\right)
:=\frac{1}{m} \sum_{i\in N} u^0(p_i)\left(
\begin{array}{c}
x^1_i\\
x^2_i
\end{array}
\right)
\]
We define the 2nd moment $M$ as the $2 \times 2$ matrix with components
\[
 M_{jk}:=\frac{1}{m} \sum_{i\in N} u^0(p_i) (x_i^j-mc^j)(x_i^k-mc^k)
\]
The eigenvalues $\lambda_1 \geq \lambda_2$ of $M$ indicate the length
and the thickness of the object.

The \emph{eccentricity} of $R$

\[
ecc:=\frac{\lambda_1-\lambda_2}{\lambda_1}=1-\frac{\lambda_2}{\lambda_1}
\]
is a measure of elongation of $R$ \citep[]{JAEHNE}.
We identify arcs as thin and elongated objects, which in mathematical
terms requires that
\[ecc \geq c_{\rm ecc}\in[0,1] \mbox{ and }
  \lambda_2 \leq c_{\rm thick}\;.\]
for given thresholds $c_{\rm ecc}$ (eccentricity threshold and $c_{\rm thick}$ (thickness threshold).\\

The detection and selection of arcs is controlled by the
three parameters $c_{\rm min2}$, $c_{\rm ecc}$ and $c_{\rm thick}$:
\begin{itemize}
\item $c_{\rm min2}$ (minimal intensity difference from background)
      is used to select arcs influenced by bright structures
      such as galaxies and dominating stars, thus it becomes active only
      locally. Choosing $c_{\rm min2}=c_{\rm min}$ is in general adequate.
\item $c_{\rm ecc}\in [0,1]$ and $c_{\rm thick}$ 
      are statistical parameters of the
      shape of an object, they are used to select elongated and thin
      structures. Choosing $c_{\rm ecc} = 0$ allows the selection of spheres 
      with a maximal 
      diameter related to $c_{\rm thick}$.
\end{itemize}

It has to be taken into account, that an object may contain several local maxima
and the region growing procedure detects adjacent parts of this object.
These parts can be merged into one object after the selection process easily .
Note that our sequence of selection and merging 
prevents objects of different shape from being merged.\\

In most cases the criteria described above are sufficient to detect arcs. 
The selection process can be refined by incorporating a priori
information, like for instance information on the center of mass
of a gravitational lens (galaxy cluster). For a spherically symmetric
 ideal
gravitational lens with center $gc=(gc_1,gc_2)$, the arcs occur
tangentially around the center. 
Let $mc$ denote the center of mass
of an arc. Then the vectors $p=mc-gc$ (position relative to center) 
and $v_2^\mu$ (approx. direction of elongation) have to be
orthogonal, giving an additional selection criterion for arcs.
However, note that user interaction is required to incorporate
a priori information on the position of the gravitational lens.

The algorithm might be also used for detection of strings, if additional post-processing steps are applied
using alternative selection criteria which take into account 
alignment information (instead of shape information as above).

\section{Results}
\subsection{Quality of detections}
The quality of the results provided by our algorithm depends mostly 
on noise variance present in the data.

Arcs may not be detected (false negative detection) by the algorithm if their intensity is within the scale of the noise. As we have seen in 
Sect.~\ref{subsec:HistogramMod} the intensity range of the arcs is in general close to the background intensity. 
The choice of parameter $c_{\rm min}$ determines whether a weak structure is interpreted as background noise (ignored) or is segmented (feasible local maxima).
Corresponding the choice of a high value for $c_{\rm min}$ increases the risk of a 
false negative detection.

Another case leading to a false negative detection is the joining of close-by objects.\\

A false positive detection may occur if the noise level exceeds  the intensity $c_{\rm min}$. In such case the edge enhancing anisotropic diffusion may recover structures present in the noise, which may be segmented as elongated objects afterwards.
Thus the choice of a low value for $c_{\rm min}$ increases the risk of a 
false positive detection. 

In the neighborhood of bright sources the background intensity increases and 
the noise may lead to several local maxima above $c_{\rm min}$.
As a result the risk of a false positive detection increases in these areas.
To reduce these kinds of false positive detections we utilize the parameter $c_{\rm min2}$ prescribing the minimal intensity difference of a detection in comparison with its surrounding. Again the quality of detection depends on the adaption of $c_{\rm min2}$ to the noise variance.\\

\subsection{Test images}

To highlight the properties of our algorithm we applied the algorithm
to three astronomical test images.
The first data set is an image of size $2285 \times 2388$ pixels with intensity range
$[-8.49,700.49]$, the second and third are of size $2048 \times 2048$
pixels with intensity ranges
$[0,19559.8]$ and $[0,9314.26]$, respectively.
We plotted in Fig.~\ref{FigHist1} the histogram of the first data set, 
the histograms of the second and third test image look similar.

\begin{figure}[h]\begin{center}
\includegraphics[width=0.45\textwidth]{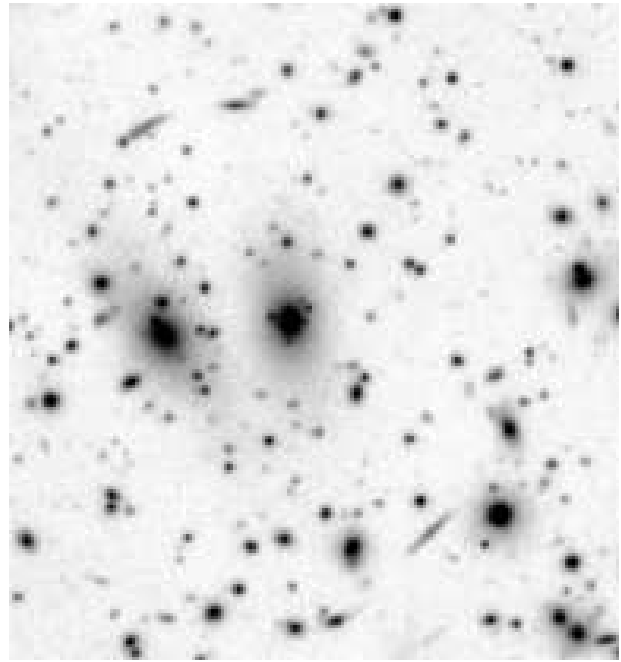}
\caption{\label{FigOriginal1} 
Detail of a VLT observation of the galaxy cluster RXJ1347-1145 (from
Miralles et al., in prep.). The
image has a size of 2285 $\times$ 2388 pixels.
This
is the first test image.}
\end{center}\end{figure}

\begin{figure}[h]\begin{center}
\includegraphics[width=0.45\textwidth]{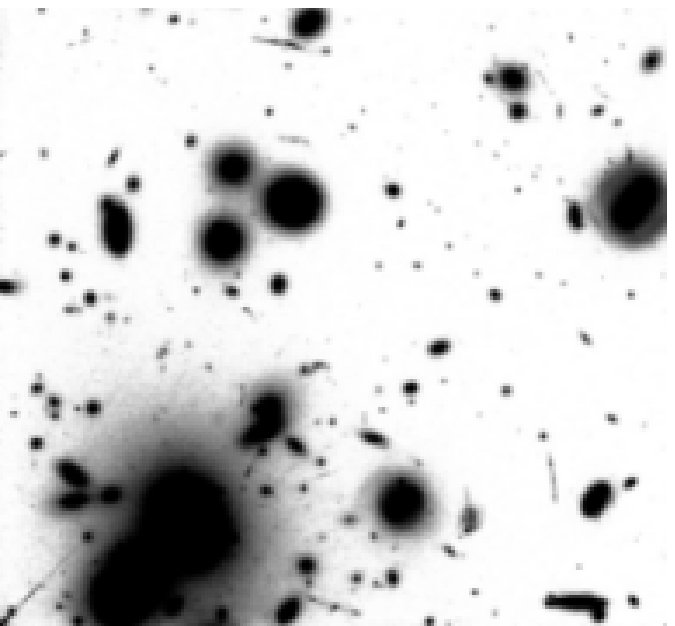}
\caption{\label{FigOriginal2} 
Detail of an HST observation of the galaxy cluster A1689. The
image has a size of 2048 $\times$ 2048 pixels. This is the
second test image.}
\end{center}\end{figure}

\begin{figure}[h]\begin{center}
\includegraphics[width=0.45\textwidth]{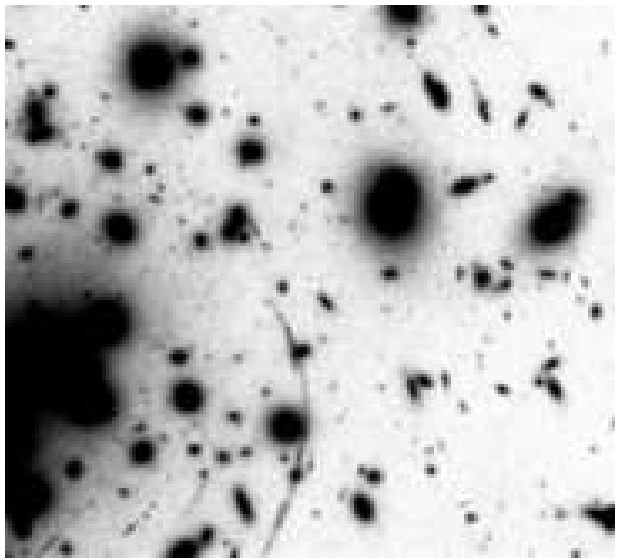}
\caption{\label{FigOriginal3} 
Detail of an HST observation of the galaxy cluster A1689. The
image has a size of 2048 $\times$ 2048 pixels. This is the
third test image.}
\end{center}\end{figure}

Figs. \ref{FigOriginal1}, \ref{FigOriginal2} and \ref{FigOriginal3}
show galaxy clusters with gravitationally lensed arcs, which have been
treated with histogram transformation; we have used
$s(x)=\sqrt{x}$ and $(a,b)=(0,1)$, $(149,200)$ and  $(141,170)$, respectively.
The histogram modification is already useful to visualize arcs.

\subsection{Computational effort}
\begin{figure}[h]\begin{center}

\includegraphics[width=0.45\textwidth]{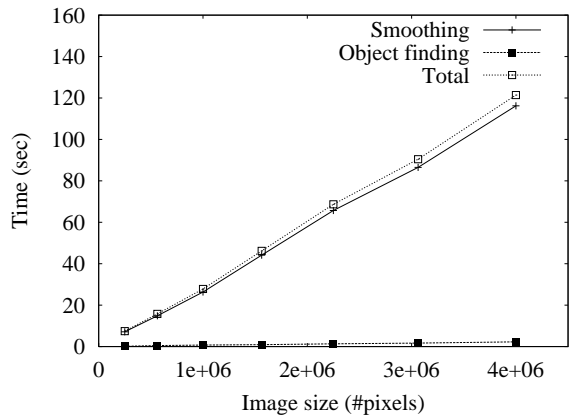}
\caption{\label{FigTime} Computation time versus 
image size for the different steps of the algorithm.}
\end{center}\end{figure}

Fig.~\ref{FigTime} shows the computation 
time\footnote{Computed on an Intel Pentium IV with 1.5 GHz} for
anisotropic diffusion filtering, object finding (including the search for local maxima, segmentation and selection) and the total computational time
in dependence of the size of the image data (number of pixels)
 in typical astronomical data sets.\\

Numerically, the pre-filtering step is most expensive as a large 
system of linear equations has to be solved.
The number of iterations performed by the conjugate gradient solver
increases slowly with growing data size. Thus  the computational effort is 
approximately linearly correlated to the number of pixels (cf. Fig.~\ref{FigTime}). 

During the detection of local maxima each pixel is invoked a fix number of times:
\begin{enumerate}
\item
to check if it is already marked,
\item
to check if it has to be 
included into a plateau during the region growing process
(once for each of its eight neighbors),
\item
to mark it as being (or not being) a local maximum.
\end{enumerate}
Therefore the computational effort for detecting the local maxima is 
linearly depending on the number of pixels.
As the segmented area is small in comparison with the image size, the effort for object finding strongly depends on the number of objects resp. 
local maxima. The same holds for the selection process.

Overall the total computation effort grows  linearly with the image size
(cf. Fig.~\ref{FigTime}).

\subsection{Anisotropic Diffusion Filtering}

\begin{figure}[h]\begin{center}
\includegraphics[width=0.45\textwidth]{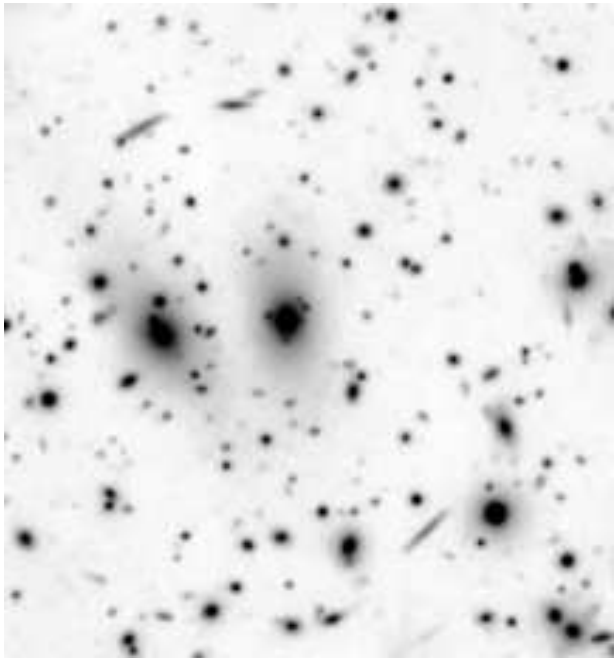}
\caption{\label{FigSmoothed1} Image of the cluster RXJ1347-1145 with
anisotropic diffusion filtering applied. Compared to the unsmoothed
image in Fig.~\ref{FigOriginal1} the noise is considerably reduced.
Parameter setting for the filtering: Filter parameter $T=15$, edge sensitivity $K=0.0001$, pre-filter
 parameter $\sigma=2$ and pre-filter parameter for structure tensor $\mu=9$}
\end{center}\end{figure}

\begin{figure}[h]\begin{center}
\includegraphics[width=0.45\textwidth]{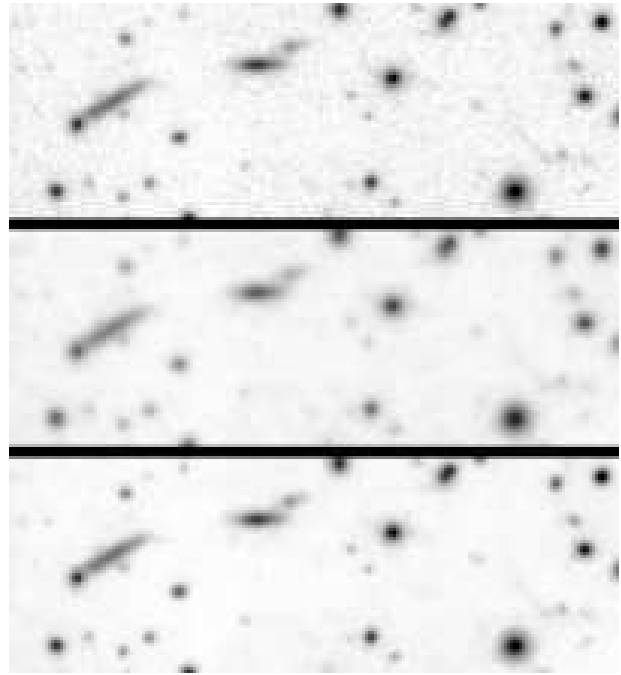}
\caption{\label{FigSmallSmoothed} Zoom of Fig.~\ref{FigOriginal1} \sl{Top}: histogram
modified data, \sl{middle}: Gaussian filtered image (Kernel $K_\sigma$
with $\sigma=5$), \sl{bottom}: image filtered with anisotropic
diffusion with the parameters: filter parameter $T=15$, edge sensitivity $K=0.0001$, pre-filter parameter $\sigma=2$ and pre-filter parameter for structure tensor $\mu=9$.
In the Gaussian filtered image (middle) the edges are not preserved well, i.e. the arcs get dispersed, while anisotropic diffusion (bottom) maintains the edges and reduces the noise at the same time.}
\end{center}\end{figure}

Fig.~\ref{FigSmoothed1} shows the result of the anisotropic diffusion
filtering for the first test data set.
To examine the effect of enhancing elongated structures we
zoom in the filtered image. Fig.~\ref{FigSmallSmoothed} shows the
histogram modified image, i.e. the cuts of the image are set to the
steep region of the distribution of the pixel values (top), the
Gaussian filtered image (middle) and the image filtered with
anisotropic diffusion (bottom). Both filters were applied separately to the 
histogram modified image.\\

The filter parameters were chosen to remove noise up to nearly the same 
signal-to-noise ratio
\footnote{
defined as SNR=$\frac{\int_\Omega \left(u^0(x)\right)^2\;dx}{\int_{\Omega}\left(u^0(x)-u(T,x)\right)^2\;dx}$
with $u^0$ the original data set and $u(T,.)$ the filtered image using Gaussian convolution
resp. anisotropic diffusion.
}
, which is about 6.1 \% in the Gaussian filtered image 
and about 6.3 \% in the image filtered with anisotropic diffusion.

In comparison to Gaussian convolution, anisotropic filtering is able to preserve accurately the edges of the objects both of high and of weak intensities. Anisotropic diffusion therefore is ideal as preprocessing for object finding based on edge detection.\\

\subsection{Detected objects}
In the first test image (Fig.~\ref{FigOriginal1}) after applying the histogram 
modification about four arcs can be recognized at first glance.
These arcs are grouped around a center of the cluster of galaxies,
which appears in the middle of the image.\\

\begin{figure}
\includegraphics[width=8cm]{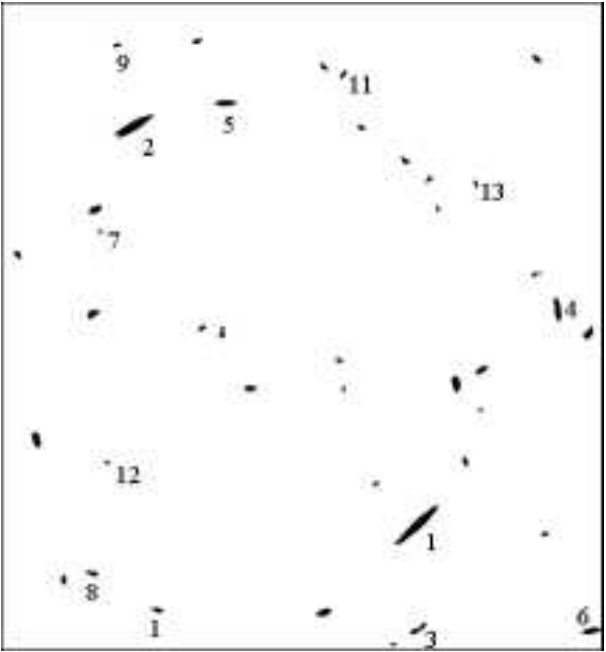}
\caption{\label{FigResult1} Result of segmentation and
selection applied to the first test image -- cluster RXJ1347-1145. Only the objects which are detected as being arcs with high probability are shown. Settings:
 intensity threshold for detection $c_{\rm min}=0.1$, minimal intensity difference from background $ c_{\rm min2}=0.1$,
eccentricity threshold  $c_{\rm ecc}=0.7$ and thickness threshold $c_{\rm thick}=16$.
}
\end{figure}

In Fig.~\ref{FigResult1} shows selected objects with
thickness $\lambda_2\le 16$ and eccentricity $ecc \ge 0.7$.
The eccentricity is color coded:
green, yellow, and red corresponds to an eccentricity in the ranges
$[0.7, 0.8]$, $[0.8,0.9]$ and $[0.9,1]$. The higher the eccentricity
and the smaller the thickness are, 
the higher is the probability that the detected object is an arc.
Incorporating a priori information on the center of the
gravitational lens, unreasonable arcs candidates can be filtered
out further (Fig.~\ref{FigCenter1}).
Table \ref{TableObjects} lists the coordinates of the detected objects with
an eccentricity greater than or equal to $0.84$ (referring to 
Fig.~\ref{FigResult1}).
Besides the arcs already mentioned the algorithm detects a significant amount of arc candidates
which are not obviously recognizable to the naked eye.

Figs. \ref{FigResult2} and  \ref{FigResult3} show the results of our algorithm applied
to the second and third test image.

\begin{table}
\centering
\caption{\label{TableObjects}List of objects detected in the first
data set (cf. Figs. \ref{FigOriginal1} and \ref{FigResult1}) with
eccentricity larger than 0.84, i.e. good arc candidates. Objects with an eccentricity larger than 0.9, i.e. objects 1 to 6 are very good candidates.}
\vspace{0,5cm}

\begin{tabular}{l|l|l|l}
&Mass center& Size (Pixel) & Eccentricity\\\hline
1  & (593,749) & 768 & 0.978\\
2  & (188,177) & 603 & 0.956\\ 
3  & (595,899) & 143 & 0.943\\ 
4  & (794,440) & 241 & 0.942\\ 
5  & (320,144) & 196 & 0.928\\
6  & (842,901) & 181 & 0.901\\
7  & (141,328) &  27 & 0.876\\
8  & (130,819) &  95 & 0.871\\
9  & (165, 61) &  52 & 0.866\\
10 & (223,871) &  86 & 0.864\\ 
11 & (488,103) &  65 & 0.854\\
12 & (151,659) &  32 & 0.845\\ 
13 & (677,260) &  44 & 0.840\\
\end{tabular}
\end{table}

\begin{table}
\centering
\caption{\label{TableParameters} Parameter settings used for the
 three test data sets.}
\vspace{0,5cm}
\begin{tabular}{l|l|l|l|l}
Param.    &  Meaning                         & Image 1 & Image 2  & Image 3\\\hline
$a$          &  low cut                         & 0      &  149     &    141\\
$b$          &  high cut                        & 1      &  200     &    170\\
$c_{\rm min}$    &  intensity threshold        	& 0.1    &  0.1     &    0.3\\
&for detection	&&&\\
$T$          &  filter parameter                & 15     &  30      &    20\\
$K$          &  edge sensitivity                & 0.0001 &  0.0001  &    0.0001\\
$\sigma$     &  pre-filter parameter             & 2      &  2       &    2\\
$\mu$        &  pre-filter parameter             & 9      &  30      &    15\\ 
&for structure tensor&&&\\ 
$c_{\rm min2}$   &  minimal intensity    &0.1     &  0.2     &    0.2\\  
& diff. from backgr. &&&\\
$c_{\rm ecc}$    &  eccentricity threshold          & 0.7    &  0.7     &    0.7\\
$c_{\rm thick}$  &  thickness threshold             & 16     &  9       &    9\\
\end{tabular}
\end{table}

\section{Comparison with software for source extraction}

In this section we compare our algorithm with the software package
``{\tt SExtractor}'' by E. Bertin \citep{BERTIN94,BERTIN96}.
{\tt SExtractor} is a general purpose astronomical software for the extraction of sources
such 
as stars and galaxies, while our software is particularly designed to 
extract gravitationally lensed arcs. 
Although the main areas of applications are rather different, several 
levels of implementation are similar, although quite different 
in details:
{\tt SExtractor} uses background estimation which in our program 
is performed by histogram modification. For \emph{Detection}, 
{\tt SExtractor} used Gaussian convolution filtering; this step 
relates to the object finding process.
\emph{Deblending} and \emph{filtering of deblending} are related to our 
merging strategy described at the second last paragraph  of the last section.\\

\begin{figure*}
\centering
\begin{tabular}{ll} 

\begin{minipage}[t]{8cm}
\includegraphics[width=8cm]{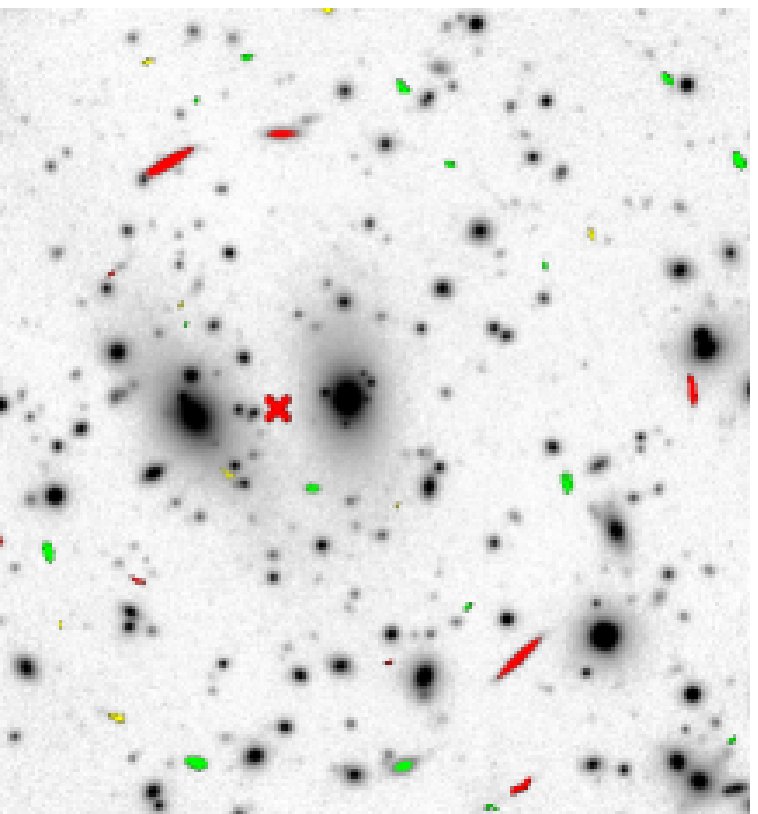}
\caption{\label{FigCenter1} Result of segmentation and
selection by taking into account  a priori information on the 
center of the gravitational lens. 
The colors encode the objects eccentricity: $[0.7,0.8]$ - green,
$[0.8,0.9]$ - yellow, $[0.9,1.0]$ - red, i.e. the red objects are most likely
arcs. The assumed 
center of the cluster is marked with a red cross.}
\end{minipage}
&
\begin{minipage}[t]{8cm}
\includegraphics[width=8cm]{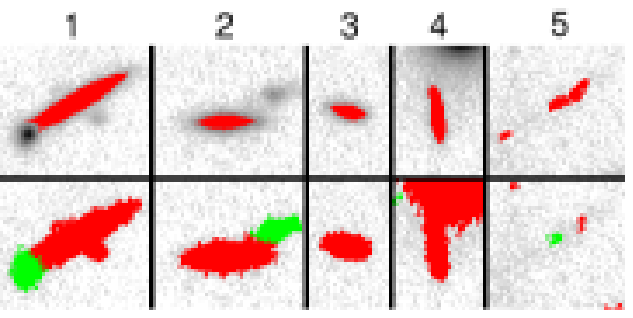}
\caption{\label{FigFiveObjects} To illustrate the features of the new algorithm we compare some specific objects
 detected by both algorithms, the proposed algorithm (first row) and the segmentation by {\tt SExtractor} (second row). The images of each column show the same part of the image. The pixels detected by the according algorithm as belonging to the arc are plotted in red. In order to distinguish between close-by objects we use green color in addition. In column 1 and 4 {\tt SExtractor} has connected pixels to the arc which do actually come from various other sources. Therefore the arc may not satisfy the required elongation and hence may not be selected. Column 5 shows that the new algorithm has detected many more pixels of the faint structure than {\tt SExtractor}.
}
\end{minipage}\\
&\\
\begin{minipage}[t]{8cm}
\includegraphics[width=8cm]{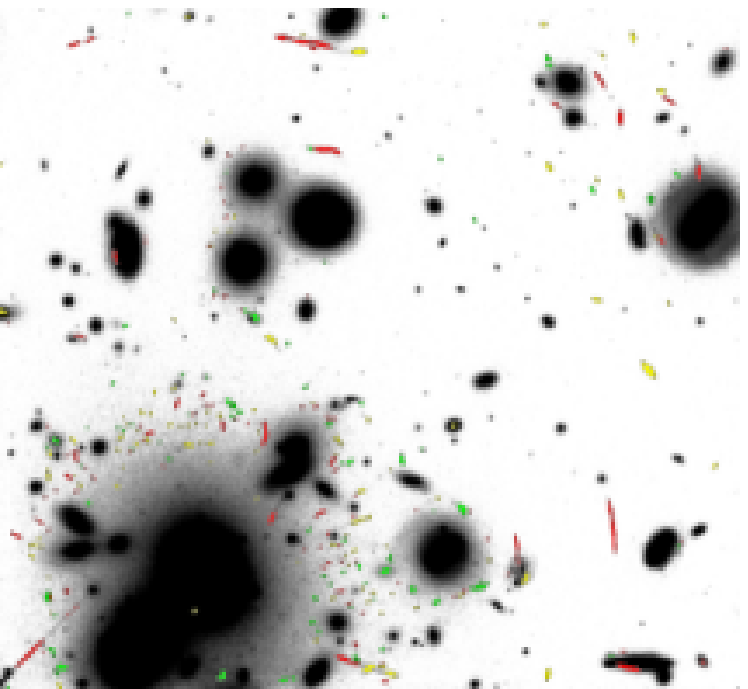}
\caption{\label{FigResult2} Result of segmentation and
selection applied to the second test image.
Parameter settings: Diffusion:
filter parameter $T=30$, edge sensitivity $K=0.0001$, pre-filter parameter $\sigma=2$ and pre-filter parameter for structure tensor $\mu=30$;
Selection:
intensity threshold for detection $c_{\rm min}=0.1$, minimal intensity difference from background $c_{\rm min2}=0.2$,
eccentricity thres\-hold  $c_{\rm ecc}=0.7$ and thickness threshold $c_{\rm thick}=9$. 
Colors as in Fig.~\ref{FigCenter1}.}
\end{minipage}
&
\begin{minipage}[t]{8cm}
\includegraphics[width=8cm]{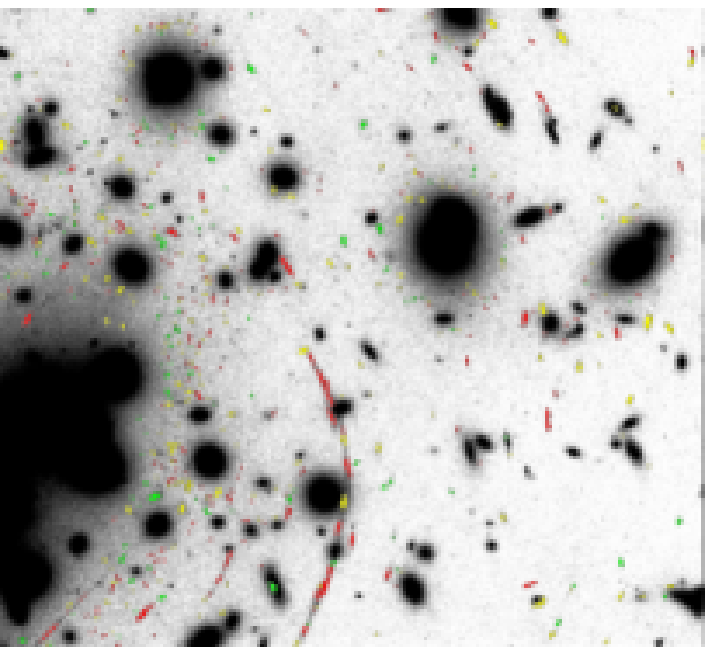}
\caption{\label{FigResult3} Result of segmentation and
selection applied to the third test image.
Parameter settings: Diffusion:
filter parameter $T=20$, edge sensitivity  $K=0.0001$,pre-filter parameter   $\sigma=2$ and pre-filter parameter for structure tensor $\mu=15$;
Selection:
intensity threshold for detection $c_{\rm min}=0.3$, minimal intensity difference from background $c_{\rm min2}=0.2$, eccentricity thres\-hold $c_{\rm ecc}=0.7$ and thickness threshold $c_{\rm thick}=9$. 
Colors as in Fig.~\ref{FigCenter1}.}
\end{minipage}
\end{tabular}
\end{figure*}

To compare both algorithms we single out five specific arcs in the first test image.
Fig.~ \ref{FigFiveObjects} shows these objects as they are detected by the 
proposed algorithm (upper row) and as they are segmented by {\tt SExtractor} (lower row).

Concerning the {\tt SExtractor's} results  a possible adaption of the {\tt SExtractor's} algorithm for detection of arcs would be to perform a selection process afterwards as described in Sect.~\ref{subsubsec:detection}. However we did not apply such a selection process
to the {\tt SExtractor's} output. Beside the arcs under investigation several other objects with only small elongations show up when applying {\tt SExtractor}. To distinguish close-by objects we use two different colors.

The first remarkable difference is that {\tt SExtractor} 
in order to measure the objects total magnitudes 
uses a far lower segmentation threshold  and thus it segments larger parts of bright objects  than our algorithm as can be realized in columns one to three in Fig.~\ref{FigFiveObjects}. The evaluation of the objects elongation then depends on the 
segmented shape.\\
The results of our algorithm show a more regular border due to the edge parallel diffusion and the edge enhancement.
Using {\tt SExtractor} one may choose a higher threshold for detection ({\tt DETECT\_THRES}) and yield a better shape of the segmented objects. However, since this is a global threshold  {\tt SExtractor} tends to loose fainter objects. Our algorithm overcomes this problem by using a local adaptive threshold.

The forth arc  in Fig.~\ref{FigFiveObjects} segmented by {\tt SExtractor} is an example for a failure of the deblending procedure. The arc is not separated from the nearby galaxy and since the composition of both sources is not much elongated it would be refused by the selection process. On the other hand tuning the parameter for deblending, which is supplied by {\tt SExtractor} for this specific arc allows a separate detection of both sources but leads to an undesired splitting of other objects.

In the fifth arc our algorithm reveals a far larger part of the weak structure than {\tt SExtractor} due to the use of anisotropic diffusion and detects its direction of elongation correctly. {\tt SExtractor} does also find an elongated part
of this arc but the direction of elongation of this part differs from the exact direction. Thus our algorithm provides a better quality of detection.

To summarize  the new algorithm  for detecting arcs presented here has 
two main advantages:

The method of filtering, e.g. anisotropic diffusion is chosen 
with regard to the kind of objects to be detected. 
The filtering process provides a closure of gaps in between objects 
as well as edge enhancing.

The use of object dependent thresholds based on edge 
detection leads to an improved segmentation even of weak sources.
A deblending procedure is not required.
Moreover the new algorithm is able to detect
close-by objects of different scale, 
for which the {\tt SExtractor's} deblending procedure fails.\\

\section{Summary and Conclusions}
We proposed a new algorithm for the detection of gravitationally lensed arcs on CCD images of
galaxy clusters, 
performing an edge-based object detection 
on the filtered image 
together with an automatic selection of arcs.\\
The algorithm consists of several steps: 
\begin{enumerate}
\item
	Histogram modification: 
	We focus on the typical range of the arcs intensities.
\item
	Filtering: we use 
	anisotropic diffusion, which enhances thin and elongated objects such 
	as arcs.
\item
	Segmentation: the principal directions as determined in the 
	diffusion process are exploited in the later segmentation
	process and the edge enhancing 
	eases border allocation. 
 \item
	Selection: after detection thin and elongated objects are selected.
\end{enumerate}

Comparing our algorithm with the software package
``{\tt SExtractor}'' we find that both algorithms rely on
\emph{multi-level} strategies for feature extraction in
astronomical data. The emphasis in {\tt SExtractor} lies on the
extraction of stars and galaxies while our algorithm is designed
to extract elongated arcs. 
Both multi-level strategies are implemented rather differently; 
the major differences are the filtering methods used and the 
determination of the segmentation threshold: {\tt SExtractor} relies on 
Gaussian convolution filtering; the threshold for segmentation depends 
on the estimated background. 
Our algorithm relies on anisotropic diffusion filtering; 
the segmentation threshold is determined by border allocation 
using the same principal directions as in the diffusion process and 
avoiding the effect of merging close-by objects. 

The new algorithm is particularly well suited for 
the detection of arcs in astronomical images. 
It can be both applied to automated surveys 
as well as to  individual clusters.

The algorithm (implemented in C) will be provided to the public.
Feel free to contact Frank Lenzen (Frank.Lenzen@uibk.ac.at).\\

\begin{acknowledgements} 
We thank Thomas Erben and Peter Schneider for kindly providing the HST
image of A1689. We are grateful to Joachim Wambsganss and Thomas Erben
for valuable comments to the manuscript.\\
The work of F.L. is supported by the Tiroler Zukunftsstiftung.
The work of O.S. and S.S. is partly supported by the Austrian Science
Foundation,  
Projects Y-123INF and P15868, respectively. 
\end{acknowledgements}

\bibliography{h4662}
\bibliographystyle{aa}

\end{document}